\documentclass[aps,prb,twocolumn,superscriptaddress]{revtex4}

\usepackage{amsmath}
\usepackage{amsfonts}
\usepackage{amssymb}
\usepackage{graphicx}
\begin{document}


\title{Mechanical, Electrical, and Magnetic Properties of Ni Nanocontacts}


\author{M. R. Calvo}
\affiliation{Departamento de Fisica Aplicada, Facultad de Ciencias, Universidad de Alicante, San Vicente del Raspeig, E-03790 Alicante, Spain}

\author{M. J. Caturla}
\affiliation{Departamento de Fisica Aplicada, Facultad de Ciencias, Universidad de Alicante, San Vicente del Raspeig, E-03790 Alicante, Spain}

\author{D. Jacob}
\affiliation{Departamento de Fisica Aplicada, Facultad de Ciencias, Universidad de Alicante, San Vicente del Raspeig, E-03790 Alicante, Spain}

\author{Carlos Untiedt}
\affiliation{Departamento de Fisica Aplicada, Facultad de Ciencias, Universidad de Alicante, San Vicente del Raspeig, E-03790 Alicante, Spain}

\author{J. J. Palacios}
\affiliation{Departamento de Fisica Aplicada, Facultad de Ciencias, Universidad de Alicante, San Vicente del Raspeig, E-03790 Alicante, Spain}



\begin{abstract}
The dynamic deformation upon stretching of Ni
nanowires as those formed with mechanically controllable break
junctions or with a scanning tunneling microscope is studied both
experimentally and theoretically. Molecular dynamics simulations
of the breaking process are performed. In addition, and in order
to compare with experiments, we also compute the transport properties
in the last stages before failure using the first-principles implementation
of Landauer$'$s formalism included in our transport
package ALACANT.
\end{abstract}


\maketitle

\section{Introduction}
In a foreseeable future, the functionality of electronic
devices will rely on the conduction properties of molecules
or nanoscopic regions comprised of a surprisingly small number
of atoms. Over the past 10 years, various experimental groups
have developed different techniques to connect two large metallic
electrodes by just an atom or a chain of atoms \cite{uno,dos,tres}. These
systems receive names such as atomic-size contacts or nanocontacts.
Although they are not expected to be of any practical
technological application in the near future, these systems are
an excellent test bed to learn about electrical transport at the
atomic scale.

While a large amount of experimental and theoretical work
has been reported for many metals, a deep theoretical understanding
is still lacking in the case of magnetic nanocontacts,
which exhibit a very rich and complex behavior \cite{cuatro}. Modeling
their mechanical, electrical, and magnetic properties with
accuracy is a challenge from which we expect to learn important
lessons on our way toward reliable theoretical descriptions
of more sophisticated systems of relevance in present and future
spin-based devices. We present here a comparison between
theoretical results of the mechanical, magnetic, and conduction
properties of Ni nanocontacts, and experiments carried out in
our laboratory.

\begin{figure}[h]
\includegraphics[width=0.9\linewidth]{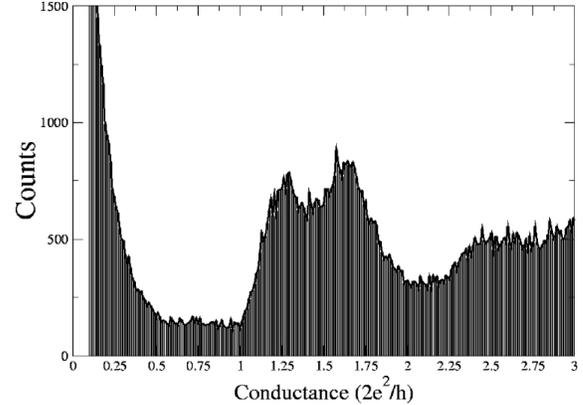}
\caption{Experimental conductance histogram for Ni nanocontacs recorded at a
bias voltage of 100 mV and a temperature of 4.2K where two low-conductance
peaks are clearly visible.}
\end{figure}

\section{Scanning Tunneling Microscope Histograms}

For the experiments, we used a high-stability scanning tunneling
microscope (STM) at low temperatures (4.2K) and under
cryogenic vacuum conditions. For both tip and sample, we used
Ni wire 99.99+\% pure. The wire was cleaned by sonication
in an acetone bath and scratched to remove contamination attached
to the surface. The conductance properties of the contacts
formed in between tip and sample were measured in a typical
two-probe configuration. A constant bias voltage (typically 10-100 mV) was applied between tip and sample, and the current
was measured using a homemade current amplifier in the range
of tens of microammeters.
\begin{figure}[h]
\includegraphics[width=0.9\linewidth]{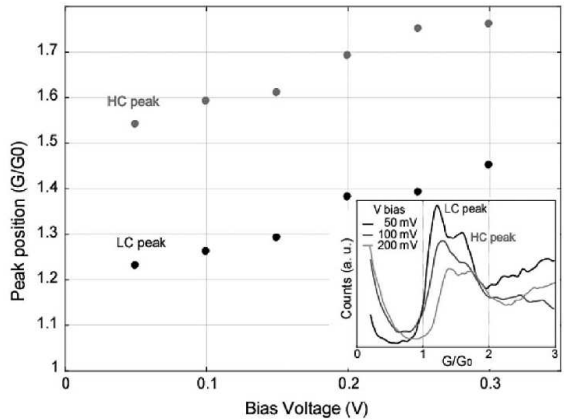}
\caption{Inset: experimental breaking histograms recorded at 4.2K at different
bias voltages. The peak above 1 $G_{0}$ can be resolved as two different peaks,
marked as lower conductance peak (LC) and higher conductance peak (HC).
The figure shows the dependence of the position of the two peaks with the
applied bias voltage averaged over different experimental realizations.}
\end{figure}
We recorded traces of conductance as a function of the relative
tip-sample distance as the two electrodes were brought together
and separated. In every trace, we made a deep tip-sample indentation
in order to prevent the repetition of the same atomistic
configurations and to assure the clearness of the contacts. Afterward,
the traces were collected to build a conductance histogram
such as the one shown in Fig. 1.

Histograms, similar to the aforementioned, for the first stages
of conduction in Ni nanocontacts have been studied before \cite{tres,cinco,seis}. There a broad peak around 1.6 $G_0$,
 where $G_0 = 2e^2/h$ is the quantum of conductance, has been reported as the first
peak after vacuum tunneling. This peak is attributed to the cases
in which the contact consists of a single atom. On the other
hand, for values of conductance below the one-atom peak, we
can notice a large amount of data coming from tunneling. This
effect is stronger in Ni than in other metals, such as Au, since for
Ni, some of the traces show a smooth transition from tunneling
to contact without a jump \cite{siete}.

Here, we have studied in detail the lowest conductance peak
after tunneling and noticed that, indeed, it is not a single
broad peak, but the superposition of two at around 1.2 $G_0$ and
1.5 $G_0$ . The position of these peaks slightly changes for different
contacts, and this may be the reason why they have not always
been clearly resolved. We have performed separated conductance
histograms from the traces for the cases of either forming
or breaking the contacts. There we find a different ratio in the
height of the peaks, being in the case of breaking traces the peak
at 1.2 $G_0$ , in general, higher than the one at 1.5 $G_0$ and vice
versa in the case of making the contacts. Finally, we notice a
dependence of the position of the peaks with the bias voltage
with a variation of even 0.3 $G_0$ in a voltage range of 300 mV, as
shown in Fig. 2.

To the best of our knowledge, the described features have not
been previously reported for any material. Normally, the peaks
in the conductance histograms do not change as a function of
the bias voltage \cite{tres}, and are equal for the cases of breaking or
forming the contacts. In order to understand our observations,
one should first try to identify all the possible atomic configurations
that could lead to a conductance in between 1 $G_0$ and
2 $G_0$ in Ni nanocontacts, and next look for the reasons that make
the value of conductance to be so dependent on the bias voltage.
In the following discussion, we will address the first question.

\begin{figure}[h]
 \includegraphics[width=0.9\linewidth]{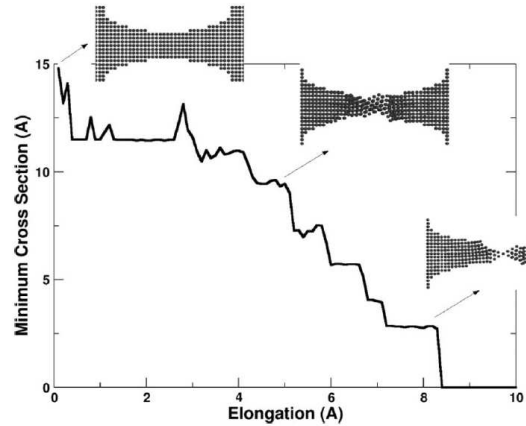}
\caption{Minimum cross section as a function of elongation along the [1 0 0]
direction for a case with 2645 atoms. Insert shows configurations at different
stages of the deformation.}
\end{figure}

\section{Mechanical Properties}

The dynamic deformation of Ni nanowires upon stretching
until failure has been studied using molecular dynamics with
empirical potentials. This type of modeling has provided significant
information about the atomic scale processes occurring
during deformation of nanowires \cite{ocho,nueve}. In this particular work,
we focus on the last stages before failure of Ni nanocontacts.
The interatomic potential for Ni developed by Mishin et al. \cite{diez}
was used in these calculations. This potential has been fitted to
reproduce the stacking fault energy of Ni. In all calculations,
deformation is achieved by displacing the outer two layers of
atoms on each side of the simulation box a fixed distance every
1000 simulation steps, similar to what is done by other
authors \cite{once}. Two different deformation velocities were used, 1
and 10 m/s \cite{once}.

The dependence of different parameters on the deformation
and, in particular, on the last stages before failure has been
studied. On one hand, we have compared the deformation of
different system sizes, between 77 and 2645 atoms, with initial
cross sections between 1.5 and 3.5$a_0$ , where $a_0$ is the lattice parameter,
$a_0$ =3.52 \AA\ , for the case where tension is applied along
the [1 0 0] direction. The dependence with crystallographic direction
has also been studied for systems with similar number
of atoms (between 610 and 658) and for directions [1 0 0],
[1 1 1], [1 1 0], and [1 1 2]. All these calculations were performed
at a fixed temperature of 4.2K by rescaling the velocities
of all atoms. Finally, the dependence with temperature was also
studied for the particular case of deformation along the [1 0 0]
direction and a cross section of 2$a_0$.

\begin{figure*}
\includegraphics[width=0.7\linewidth]{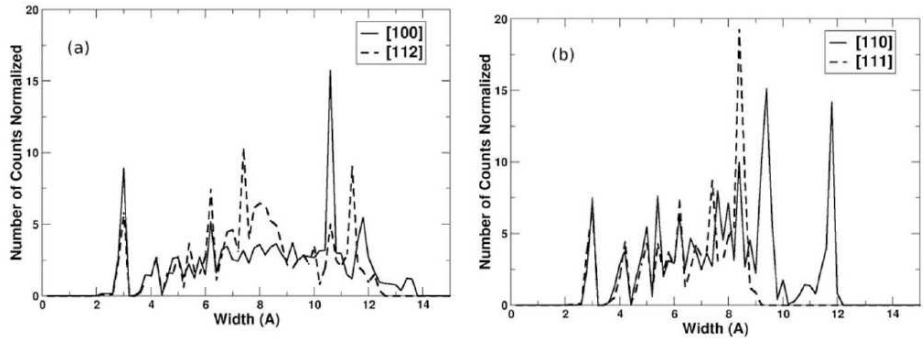}
\caption{Histogram of minimum cross sections from 100 independent simulations for crystallographic directions. 
(a) [1 0 0] and [1 1 2]. (b) [1 1 1] and [1 1 0].}
\end{figure*}

The minimum cross section perpendicular to the applied tension
is computed every 1000 steps following the method developed
by Bratkovsky et al. \cite{dieciocho}. This method allows us to
compare the results of deformation along different crystallographic
directions. Fig. 3 shows the minimum cross section as
a function of elongation obtained for one particular case with
2645 atoms and deformation along the [1 0 0] direction. The
insets show several configurations during the deformation: the
initial configuration, an intermediate configuration that is particularly
stable, and the final configuration, which, in this case,
consists of a single atom, a monomer, connecting the two sides
of the nanowire. The mechanisms for deformation at this scale
have been studied in detail by other authors \cite{once}. Consistent with
their work, we observe the sliding of planes during deformation
(such as in the intermediate inset in Fig. 3) that results in a narrowing
of the wire. This results in preferential configurations,
reflected in the plateaus observed in Fig. 3.

The dependence of the deformation on the stretching direction
has been studied for [1 0 0], [1 1 1], [1 1 0], and [1 1 2] directions.
Static calculations to obtain the energy to fracture along these
different directions show that the [1 1 2] direction has the lowest
energy per unit surface, followed by the [1 0 0], [1 1 0], and
[1 1 1] directions. However, this behavior could be different
during dynamic deformation. For each direction, calculations
were repeated 100 times in order to gain some statistics and
obtain a histogram of cross sections. Fig. 4 shows the histograms
obtained for all crystallographic directions with clear peaks at
particular cross sections. These preferential cross sections do not
depend on the system size. Histograms obtained from smaller
systems result in peaks located at exactly the same positions.
It is interesting to point out that the first four peaks appear
in all cases. These correspond to the smallest cross sections,
consisting of less than three atoms across.

\begin{figure}
 \includegraphics[width=0.9\linewidth]{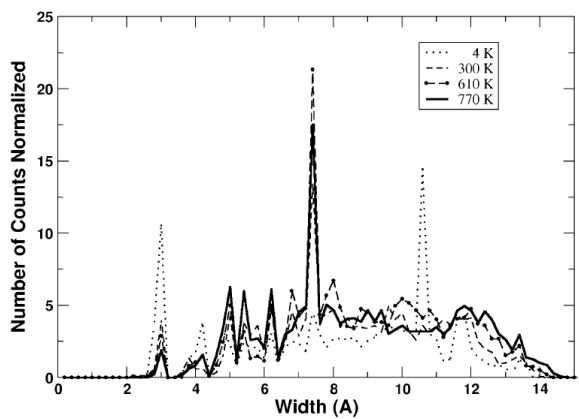}
\caption{Histogram of minimum cross sections from 25 independent simulations
and four different temperatures. [1 0 0] crystallographic direction.}
\end{figure}

These first peaks are also very stable with temperature. Fig. 5
shows the histograms obtained for four different temperatures,
4.2K as before and 300K, 610K, and 770K. These histograms
were obtained from 25 independent calculations and for the
[1 0 0] direction. Notice that from 4.2K to room temperature,
there is a strong reduction in the first peak. Therefore, the contact
breaks very rapidly at high temperatures. On the contrary, structures
with cross sections of two or three atoms seem to be very
stable with temperature. For wider structures, there is not a clear
peak at high temperature as in the case of 4.2K, which seems
to point to very different types of structures possible when temperature
increases. The dependence with temperature of these
structures has been studied previously by other authors \cite{diecisiete}.

In what follows, we focus on the final stage before failure
of these nanocontacts. Two structures have been identified: a
monomer, where a single atom acts as a bridge between the two
contacts, and a dimer,where two atoms aligned forming a bridge
between the two sides of the wire. These two configurations are
shown in the insets of Fig. 6, (a) being the monomer and (b)
the dimer. In order to identify the dimer, we have calculated
the total number of neighbors for each atom and the number of
neighbors to the left of the atom position and to the right along
the z-direction. In this manner, it is easy to identify a dimer since
it will consist of two neighboring atoms each one with only one
neighbor on one side, one atom to the left, the second one to the
right. From all the cases computed including all crystallographic
directions (4 0 0), a total of 82\% form a dimer before failure,
while only 18\% break from the monomer. In the cases studied
in detail, the monomer is formed before the dimer, but, in a few
cases, the contact breaks before forming the dimer.

\section{Transport properties}

We have finally computed the transport properties of the
two possible types of configurations, i.e., a monomer and a
dimer, before failure. The basics to calculate the zero-bias,
zero-temperature conductance, $G$, in a metallic nanocontact are
contained in Landauer$'$s formalism, where $G$ is proportional to
the quantum mechanical transmission probability of the electrons
at the Fermi energy, $E_F$

\begin{equation}
 G=\frac{e^2}{h}[T_{\uparrow}(E_{F})+T_{\downarrow}(E_F)]
\end{equation}

In this expression, the contributions from spin up (majority)
and spin down (minority) channels have been explicitly separated,
while the contribution from all the orbital channels has
been condensed in $T$. For simplicity, we assume no spin mixing
due to either spin-orbit scattering or noncollinear magnetic
structures at the bridge. The detailed electronic and magnetic
structure of the nanocontact is important, and, in order to achieve
a quantitative level of agreement with experiments, one has to
rely on first-principles or ab initio calculations. These calculations
are performed with our code ALACANT \cite{doce,trece,catorce,quince}. The
details of the calculation have been presented in previous publications
\cite{doce,trece,catorce,quince}. Essentially, one computes the self-consistent
Kohn-Sham Hamiltonian for the narrowest part of the nanocontact,
replacing the rest of atoms by a self-energy calculated based
on a parametrized Bethe lattice.

\begin{figure}
 \includegraphics[width=0.9\linewidth]{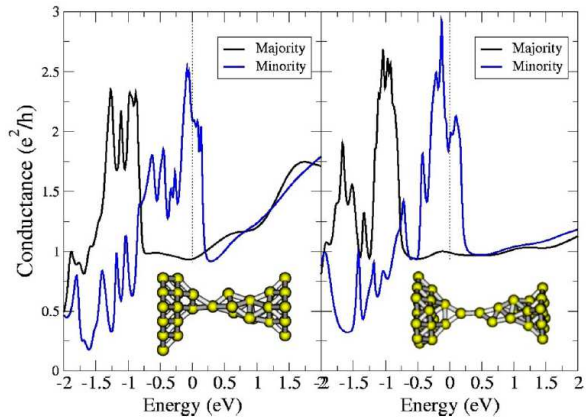}
\caption{(Left panel) Transmission for both spin species as a function of energy
for the monomer configuration (shown in the inset) before failure. (Right panel)
The same, but for a dimer configuration before failure.}
\end{figure}

Using as input data two representative atomic configurations,
as those shown in Fig. 6, the transmission spectrum of these two
structures has been calculated at the local spin density approximation
(LSDA) level, and close attention has been paid to the
choice of basis set for the central part of the nanocontact. As
expected for Ni, majority conduction is smooth as a function of
energy due to the $s$-like nature of this channel, while minority
conduction is strongly fluctuating close to the Fermi energy due
to the $d$-like character of this channel \cite{dieciseis}. Interestingly, the
average value of the conductance around the Fermi level for
both the examples lies somewhere in the vicinity of 1.6($2e^2/h$),
which agrees fairly well with the value of the highest conductance
peak in the histogram (see Fig. 1). Remarkably, this value
can only be obtained with an LSDA Kohn-Sham potential. The
use of generalized gradient corrected functionals or hybrid functionals
reduces strongly the conductance for minority electrons,
leaving no possible explanation for the high-conductance peak.

As far as the origin of the low-conductance peak, one could
possibly attribute it to the presence of a domain wall at the
narrowest section. Domain walls have a small but sizeable effect
on the conductance of Ni nanocontacs, reducing it by an
amount that agrees in magnitude with the conductance of the
lowest peak. More experimental and theoretical work is, however,
needed in this direction before this hypothesis can be
confirmed.

\begin{acknowledgments}
This work was supported in part by the Ministry of Education and Science (MEC) of
Spain under Grant MAT2007-65487, Grant FIS2004-02356, and Grant CONSOLIDER
CSD2007-00010 and in part by the Generalitat Valenciana under
Grant ACOMP06/138. The works of M. J. Caturla and C. Untiedt were supported
by the Spanish McyT. F. under a “Ram\'on y Cajal” grant.
\end{acknowledgments}

\bibliography{ieee_latex}

\end{document}